\documentclass[a4paper,fleqn,usenatbib,letters]{mnras}

\usepackage{newtxtext,newtxmath}

\usepackage[T1]{fontenc}
\usepackage{ae,aecompl}

\usepackage{graphicx}	
\usepackage{amsmath}	
\usepackage{amssymb}	
\usepackage{multirow}
\usepackage{listings}

\title[Candidates for binary systems with RR~Lyrae stars]{Candidates for RR~Lyrae in binary systems from the OGLE Galactic bulge survey}

\author[Prudil et al.]{
Z. Prudil$^{1}$\thanks{E-mail: prudilz@ari.uni-heidelberg.de}, M. Skarka$^{2,3,5}$, J. Li{\v s}ka$^{4,5}$, E. K. Grebel$^{1}$, C.-U.~Lee$^{6}$ \\
$^{1}$ Astronomisches Rechen-Institut, Zentrum f{\"u}r Astronomie der Universit{\"a}t Heidelberg, M{\"o}nchhofstr. 12-14, D-69120 Heidelberg, Germany\\
$^{2}$ Department of Theoretical Physics and Astrophysics, Masaryk University, Kotl\'{a}\v{r}sk\'{a} 2, CZ-611 37, Czech Republic\\
$^{3}$ Astronomical Institute, Czech Academy of Sciences, Fri\v{c}ova 298, CZ-251 65, Ond\v{r}ejov, Czech Republic\\
$^{4}$ Central European Institute of Technology - Brno University of Technology (CEITEC BUT), Purky\v{n}ova 656/123, CZ-612 00 Brno, \\ ~~Czech Republic\\
$^{5}$ Variable Star and Exoplanet Section of the Czech Astronomical Society, K Lesu 345/10, CZ-142 00 Prague-Kam\'{y}k, Czech Republic \\
$^{6}$ Korea Astronomy and Space Science Institute, 776 Daedeokdae-ro, Yuseong-Gu, Daejeon 34055, Republic of Korea \\
}

\date{Accepted XXX. Received YYY; in original form ZZZ}

\pubyear{2019}

\begin{document}

\label{firstpage}
\pagerange{\pageref{firstpage}--\pageref{lastpage}}
\maketitle

\begin{abstract}
We present 20 newly discovered candidates for binary systems with an RR~Lyrae companion. Using the photometric data from the Optical Gravitational Lensing Experiment (OGLE) and Korea Microlensing Telescope Network (KMTNet) for the Galactic bulge, we searched for binary candidates among non-modulated fundamental-mode RR~Lyrae variables. We constructed and investigated over 9000 $O-C$ diagrams for individual pulsators, in order to find signs of the light-travel time effect. We found 20 variables exhibiting cyclic variation in the $O-C$ (time of the observed minus calculated brightness maximum) diagram, which can be associated with a second component in the system, but for confirmation of binarity, a long-term radial velocity study is necessary. The modeling of the $O-C$ diagrams yields orbital parameters, estimates of the semi-amplitude of the radial velocity curve, and the mass function. The orbital periods of our candidates range from 3 to 15 years. Two of the binary candidates display a minimum mass higher than the assumed mass of the RR~Lyrae component, which points towards an evolved companion that can under some circumstances contribute significantly to the total light of the system. 
\end{abstract}


\section{Introduction}

Pulsating stars of RR Lyrae type (hereafter RRL) have played a prominent role since their discovery \citep{Pickering1895}. As one of the fundamental distance indicators, they help to map the Milky Way and its formation history \citep{Drake2013,Sesar2013,Hernitschek2019} and close galactic systems \citep{Haschke2012a,Haschke2012b,Jacyszyn-Dobrzeniecka2017}. 

Knowledge of the luminosity of RRL is, therefore, crucial. The luminosity of an RRL is mainly defined by the helium and heavy-element abundances and by the mass of the star. The latter one is still poorly known and can be deduced only indirectly from stellar evolutionary and pulsation models \citep{Sweigart1987,Lee1990,Popielski2000}. An independent determination of the mass would constrain the accuracy of the theoretical predictions and will have a large impact on all topics employing RRL stars. This motivates the search for RRL stars in binary systems, as binaries offer an excellent means of accurate stellar mass determinations, but very few such candidate systems have been found so far \citep{Saha1990,Soszynski2009OGLEIIILMC,Soszynski2011OGLEIII,Li2014MNRAS,Hajdu2015,Liska2016b,Kervella2019a}. Identification of binary systems will also help to answer the question why binaries with an RRL companion are so rare. Currently, we know about 150\,000 RRL stars in the Milky Way. Even if we assume that only a few percent of these RRL stars are in binary systems \citep{Hajdu2015}, we should detect a large number of binaries. The shortcomings of the detection caused by the evolutionary status of RRL stars are discussed in \cite{Skarka2016}.

Stellar dynamical mass can be derived when a star is bound in an eclipsing binary system. However, here comes the issue: to our current knowledge, no eclipsing binary containing an RRL star has been discovered and verified. There are over one hundred candidates \citep[see the online version of the database\footnote{\url{https://rrlyrbincan.physics.muni.cz/}},][]{Liska2016c}, and \cite{,Kervella2019a,Kervella2019b}, but only TU UMa is bound in a binary system with high confidence \citep{Wade1999,Liska2016a,Kervella2019a}. The binarity is mostly inferred from the cyclic period variations (mostly with years-long periods) as a consequence of the light-travel time effect\footnote{The LiTE hypothesis is based on the possible existence of an additional hidden object in the system. The RRL and the companion move around a common barycenter, and the times of brightness maxima of the pulsator's changes occur earlier or later, based on its orbital position relative to the observer.} \citep[hereafter LiTE, for example,][just to mention some studies]{Olah1978,Hajdu2015,Li2018}. A comprehensive review of the binary RRL candidates can be found in \citet{Liska2016b}.

\citet{Karczmarek2017} found that among binary stars with orbital periods below 2000\,d, mass transfer can occur and binary evolution pulsators, which only mimic RRL type stars, can be formed \citep{Pietrzynski2012,Smolec2013}. Thus, only wide systems give a good chance to reveal a real RRL in a binary system. At the same time, the large size of the semi-major axis of wide binaries decreases the chance to observe eclipses. The observation of the cyclic pulsation period variations itself cannot give an unambiguous proof of binarity. \citet{Skarka2018} showed for Z CVn that the years-long spectacular pulsation period variations can have a different origin than LiTE. This shows that some of the candidates are probably false positives.

Obtaining radial velocity measurements over several years is, therefore, highly desired \citep[e.g.,][]{Fernley1997,Solano1997}. There have been a few ongoing projects for monitoring the candidates \citep{Guggenberger2016,Poretti2018}. Probably the most promising results can be expected from \citet{Hajdu2018}, who have observed candidates from the Galactic bulge and who detected the expected differences in radial velocities for two of their sample stars in two different seasons.

In this paper, we identify 20 new binary candidates using an analysis of their pulsation period variations. We describe our sample and methods in Sect. \ref{sec:SamSel} and discuss the results in Sect. \ref{sec:Discus}. 

\section{Data selection and analysis} \label{sec:SamSel}

For the purpose of this study, we used $I$-band photometric data of RRLs from the OGLE Galactic bulge (OGLE GB) survey \citep{Soszynski2011OGLEIII,Soszynski2014OGLEIV,Soszynski2017OGLEIV}. Our search was aimed at the LiTE manifestation in the $O-C$ diagram \citep[{\it O} -- observed, {\it C} -- calculated time of the brightness maximum based on the star's ephemerides,][]{Irwin1952a,Sterken2005}, which may hint towards an unseen companion of a variable star. In order to avoid confusion with modulated Blazhko stars \citep{Blazhko1907}, we investigated only the non-modulated fundamental mode variables from studies focused on the occurrence rate of the Blazhko effect \citep{Prudil2017Blazhko,Prudil2019OO}. We did not make any additional cuts based on the data quality or quantity, and in the end we analysed more than 9000 stars, in order to increase the chances of finding candidates. We note that in the end we identified almost 200 stars with a peculiar change in the $O-C$ diagram, but for our current study, we chose only those with strong signs of the LiTE.

In the first step of the analysis, we constructed the light curve templates for individual variables based on the entire OGLE GB photometry using a Fourier decomposition of high degree $n$:
\begin{equation} \label{eq:FourierSeries}
m\left ( t \right ) = A_{0}^{I} + \sum_{k=1}^{n} A_{k}^{I} \cdot \text{cos} \left (2\pi k \vartheta + \varphi_{k}^{I} \right ),
\end{equation}
where $A_{k}^{I}$ and $\varphi_{k}^{I}$ represent amplitudes and phases, respectively, and $A_{0}^{I}$ stands for the mean magnitude. The $\vartheta$ stands for the phase function defined as:
\begin{equation} \label{eq:PhaseFunction}
\vartheta =\frac{HJD-M_{0}}{P},
\end{equation}
where the $HJD$ represents the time of observation in Heliocentric Julian Date, $M_{0}$ denotes the time of zero epoch (in our case for maximum brightness), and $P$ is the pulsation period. 

We note that the photometric data from OGLE GB phases III and IV differ by an offset in their mean magnitudes (by $\approx 0.01$\,mag). In order to compensate this effect, we shifted the OGLE III data by the difference in the mean apparent magnitudes, $A_{0}^{I}$, between the aforementioned data releases. In the second step, we binned the OGLE photometry based on the observational seasons (where each bin represents a point in the $O-C$ diagram), and in some cases, where sufficient data were available (more than 300 observations per season), we divided each season into two bins since the data were sufficient in a given observing season. 

\begin{figure}
\includegraphics[width=\columnwidth]{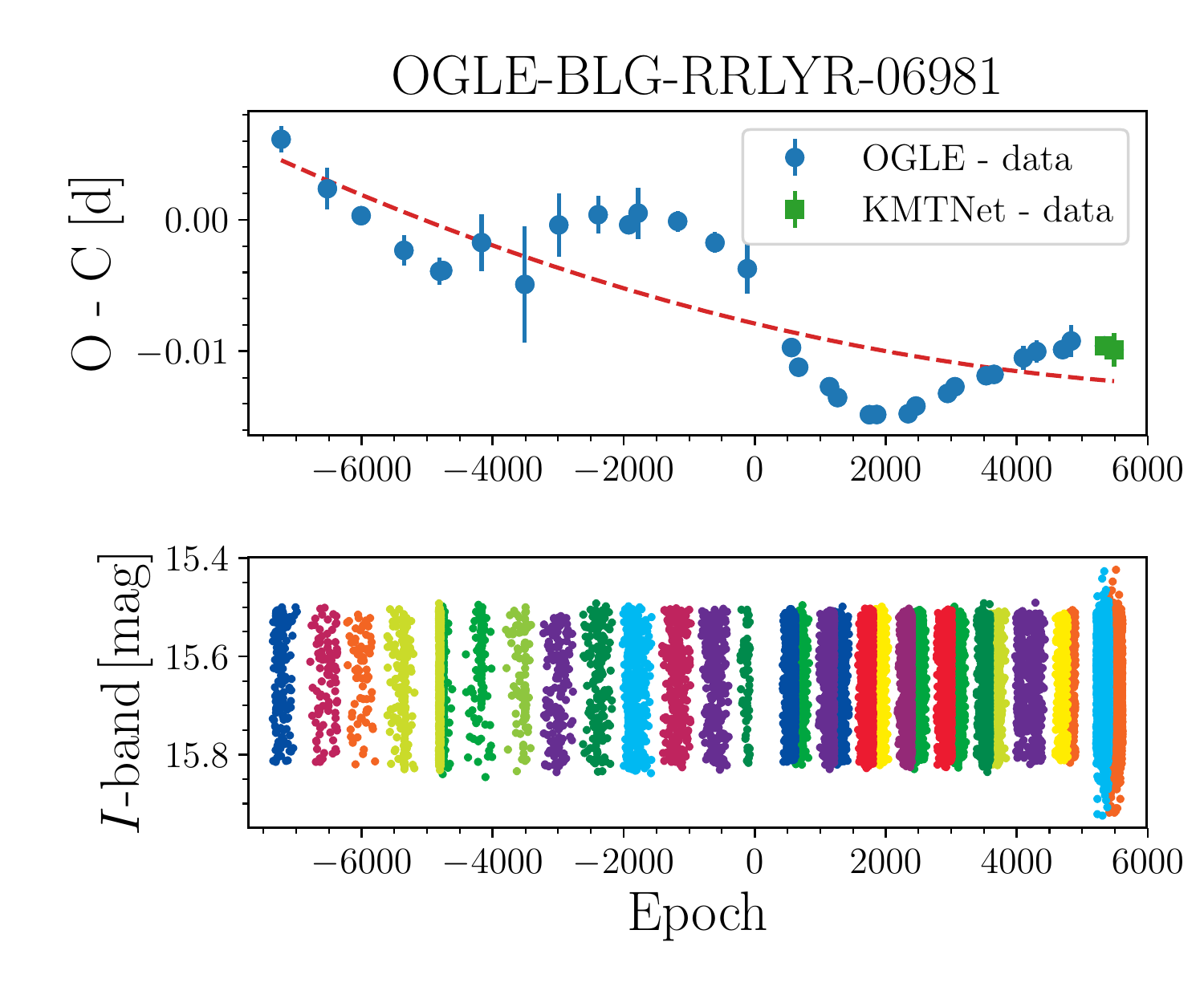}
\caption{The analysis of the $O-C$ diagram for one of the candidate stars. In the top panel, the blue points represent the time difference between the observed and calculated time of the maxima ($O-C$) and the red dashed line represents the secular variation in the pulsation period with blue points indicating data from the OGLE survey and green squares standing for the photometry from the Korea Microlensing Telescope Network \citep[KMTNet,][]{Lee2014,Kim2016}. In the bottom panel, we show the available photometric data, color-coded based on the data bins for the $O-C$ diagram.}
\label{fig:OCexample}
\end{figure}

The data in each bin were phased using the ephemerides (pulsation period $P$ and time of brightness maximum $M_{0}$) provided by the OGLE team, and fitted with a light curve template created using the whole sample. The $O-C$ was calculated for each binned phased curve as a shift in phase of the maximum brightness from the zero point, and multiplied by the pulsation period. Using a bootstrap resampling of each bin we estimated the errors of the individual points in the $O-C$ diagram. An example of the $O-C$ diagram is shown in Fig.~\ref{fig:OCexample}.

For the subsequent visual examination, each $O-C$ diagram was corrected for the changes in the pulsation period using the following equation:
\begin{equation} \label{eq:PolyChange}
O-C = a_{0} + a_{1} \cdot E + a_{2} \cdot E^{2},
\end{equation}
where $E$ represents the number of cycles that elapsed since the $M_{0}$, and $a_{0,1,2}$ represents free parameters of the parabolic fit, with $a_{2}$ denoting the pulsation period change rate. Such corrected $O-C$ diagrams were visually inspected for possible binary patterns (sinusoidal like waves). In doing so, we independently re-discovered candidates found by \citet{Hajdu2015} and \citet{Hajdu2018}, thus verifying our approach. For variables that were selected for further analysis we used data uncorrected for period variation and fitted them using the combination of Eq.~\ref{eq:PolyChange} and eq.~2 from \citet{Irwin1952a}:

\begin{align} \label{eq:Irwin}
O-C = a_{0} + a_{1} E + a_{2} E^{2} + A \left [ \left ( 1 - e^{2} \right ) \frac{\text{sin}(\nu + \omega )}{1 + e \text{cos}\,\nu} + e \text{sin}\,\omega \right ].
\end{align}
In this relation, $\nu$ is the true anomaly, $\omega$ is the argument of periastron, $e$ is the numerical eccentricity, and $A$ is the shift in radial position in light days. We refer the interested reader to the appendix~A in \citet{Liska2016a} for a thorough description of the individual parameters. 

For stars with $O-C$ diagrams that can be described by Eq.~\ref{eq:Irwin} we also further tested whether they exhibit the Blazhko modulation. We utilized the binned data and searched for an amplitude $A_{1}$ variation (derived from Eq.~\ref{eq:FourierSeries}) similar to the trend in the $O-C$ diagram. Furthermore, we tested our stars for the Blazhko effect using the method from \citet{Shibahashi2017}, but the results were inconclusive with the available data. In total, for 20 stars, which were visually selected, Eq.~\ref{eq:Irwin} is valid and we will refer to them from here on as our binary candidates. 

In order to increase the observational time span and to verify our binary candidates, we utilized data from the Korea Microlensing Telescope Network \citep[KMTNet;][]{Lee2014,Kim2016} collected in 2018. The photometry was processed in a similar manner as the OGLE dataset and incorporated in our calculated $O-C$ diagrams. The orbital parameters were then determined iteratively in the following way: the $O-C$ diagrams of the selected candidates were modeled using Eq.~\ref{eq:Irwin}, their light curves were then corrected for the LiTE effect, and in a next step served as a dataset for obtaining a light curve template, which is more accurate than the raw one. Using the new more accurate templates, we then derived new $O-C$ diagrams and subsequently the new orbital parameters.

\subsection{Orbital parameters of candidates} \label{sec:Candidates}

\begin{figure*} 
\includegraphics[width=504pt]{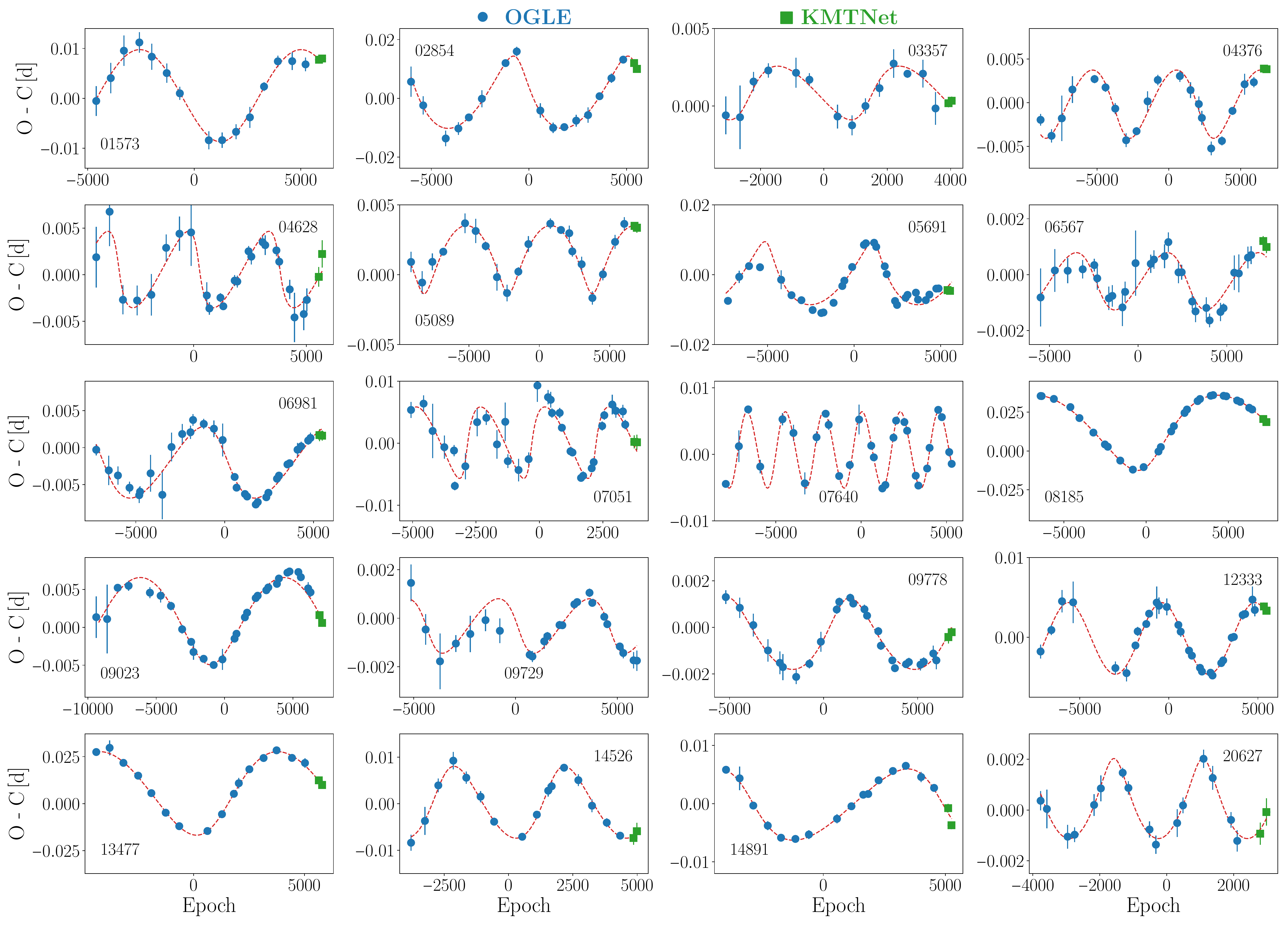}
\caption{The $O-C$ diagrams for all of the candidates with the red dashed line representing the binary model based on Eq.~\ref{eq:Irwin} with subtracted trend for the period change. The blue points stand for photometry from the OGLE survey and green squares represent data from the KMTNet. The legend for each $O-C$ diagram contains the OGLE ID in the form OGLE-BLG-RRLYR-ID.}
\label{fig:Elipsoids-cont}
\end{figure*}

In this subsection, we discuss the characteristics of 20 newly discovered binary candidates with an RRL component. In Fig.~\ref{fig:Elipsoids-cont} we show a mosaic of the $O-C$ diagrams for all of the candidates with a model based on the Eq.~\ref{eq:Irwin}. Using this model, we derived some of the orbital and physical properties of our candidate variables. In addition, we estimated the semi-amplitude of the radial velocity of RRL component $K$ in km\,s$^{-1}$ using the modified relation from \citet{Irwin1952b}:
\begin{equation}
K = \frac{2\,\pi\,a \text{sin} i \cdot \text{au}}{8.64 \cdot 10^{7} P_{\rm orbit} \sqrt{\left ( 1 - e^{2} \right )}} .
\end{equation}
where $P_{\rm orbit}$ represents the orbital period, $a$ is the semi-major axis of RRL component, au is astronomical unit, and $i$ is the inclination angle of the orbit. Furthermore, through the third Kepler law, we can calculate the mass function $f(\mathfrak{M})$ as:
\begin{equation}
f(\mathfrak{M}) = \frac{4\,\pi^{2}\,\left ( a\, \text{sin}\,i \right )^{3}}{\text{G} \cdot P^{2}_{\rm orbit}}. 
\end{equation} 
The list of candidates with their physical and orbital properties from the $O-C$ modeling can be found in Tab.~\ref{tab:BinTabulka}.

\begin{table*}
\caption{Table of derived model parameters for the binary candidates. Column 1 contains the ID of the star in the form OGLE-BLG-RRLYR-ID. Columns 2 and 3 list the orbital period and semi-major axis in au. The numerical eccentricity and the argument of periastron in degrees are listed in columns 4 and 5. The columns 5 and 6 list the semi-amplitude of the radial velocity and period change rate. The column 7 and 8 contain the mass function, and minimum mass of a possible companion for an (adopted fixed mass of the RRL component of 0.65 M$_{\odot}$).}
\label{tab:BinTabulka}
\begin{tabular}{lrrccccccc}
\hline 
\#ID & $P_{\rm orbit}$\,[d] & $a$\,sin\,$i$\,[au] & $e$ & $\omega$\,[deg] & $K$\,[km\,s$^{-1}$] & $\beta$\,[d\,Myr$^{-1}$] & $f(\mathfrak{M})$\,[M$_{\odot}$] & $M_{\rm \ast}$\,[M$_{\odot}$]\\ \hline 
01573 & 4191 $\pm$ 90 & 1.600 $\pm$ 0.102 & 0.054 $\pm$ 0.058 & 272 $\pm$ 93 & 4.16 & 0.59 $\pm$ 0.05 & 0.03108 & 0.305 \\
02854 & 3517 $\pm$ 29 & 2.199 $\pm$ 0.079 & 0.452 $\pm$ 0.021 & 122 $\pm$ 9 & 7.62 & 0.08 $\pm$ 0.02 & 0.11458 & 0.548 \\
03357 & 3163 $\pm$ 117 & 0.318 $\pm$ 0.036 & 0.356 $\pm$ 0.120 & -20 $\pm$ 27 & 1.17 & 0.40 $\pm$ 0.03 & 0.00043 & 0.060 \\
04376 & 2869 $\pm$ 43 & 0.688 $\pm$ 0.049 & 0.227 $\pm$ 0.151 & -142 $\pm$ 10 & 2.68 & 0.07 $\pm$ 0.01 & 0.00528 & 0.150 \\
04628 & 2088 $\pm$ 34 & 0.891 $\pm$ 0.086 & 0.609 $\pm$ 0.078 & 171 $\pm$ 11 & 5.86 & 0.04 $\pm$ 0.03 & 0.02165 & 0.262 \\
05089 & 2874 $\pm$ 23 & 0.422 $\pm$ 0.018 & 0.453 $\pm$ 0.053 & -91 $\pm$ 13 & 1.79 & 0.11 $\pm$ 0.01 & 0.00121 & 0.087 \\
05691 & 3682 $\pm$ 66 & 1.639 $\pm$ 0.091 & 0.585 $\pm$ 0.043 & 120 $\pm$ 13 & 5.97 & -0.10 $\pm$ 0.04 & 0.04330 & 0.352 \\
06567 & 2368 $\pm$ 57 & 0.183 $\pm$ 0.021 & 0.250 $\pm$ 0.178 & -174 $\pm$ 26 & 0.87 & 0.08 $\pm$ 0.01 & 0.00015 & 0.041 \\
06981 & 4260 $\pm$ 107 & 0.883 $\pm$ 0.056 & 0.318 $\pm$ 0.071 & 165 $\pm$ 16 & 2.38 & 0.04 $\pm$ 0.02 & 0.00505 & 0.148 \\
07051 & 2197 $\pm$ 27 & 1.200 $\pm$ 0.090 & 0.589 $\pm$ 0.009 & -19 $\pm$ 5 & 7.35 & 0.42 $\pm$ 0.04 & 0.04772 & 0.367 \\
07640 & 1243 $\pm$ 02 & 1.000 $\pm$ 0.015 & 0.109 $\pm$ 0.001 & -362 $\pm$ 3 & 8.80 & 0.05 $\pm$ 0.01 & 0.08626 & 0.479 \\
08185 & 5041 $\pm$ 22 & 4.281 $\pm$ 0.025 & 0.338 $\pm$ 0.004 & -53 $\pm$ 2 & 9.82 & -0.35 $\pm$ 0.01 & 0.41182 & 1.067 \\
09023 & 5001 $\pm$ 133 & 0.998 $\pm$ 0.045 & 0.184 $\pm$ 0.051 & -101 $\pm$ 17 & 2.21 & 0.51 $\pm$ 0.02 & 0.00530 & 0.150 \\
09729 & 2211 $\pm$ 53 & 0.208 $\pm$ 0.027 & 0.367 $\pm$ 0.197 & 194 $\pm$ 15 & 1.10 & -0.17 $\pm$ 0.01 & 0.00025 & 0.049 \\
09778 & 3177 $\pm$ 52 & 0.269 $\pm$ 0.011 & 0.292 $\pm$ 0.044 & 69 $\pm$ 2 & 0.96 & 0.12 $\pm$ 0.01 & 0.00026 & 0.050 \\
12333 & 3183 $\pm$ 25 & 0.778 $\pm$ 0.024 & 0.045 $\pm$ 0.046 & 74 $\pm$ 49 & 2.66 & 0.05 $\pm$ 0.01 & 0.00620 & 0.160 \\
13477 & 4609 $\pm$ 160 & 3.882 $\pm$ 0.231 & 0.173 $\pm$ 0.021 & -37 $\pm$ 11 & 9.30 & -0.13 $\pm$ 0.19 & 0.36721 & 1.000 \\
14526 & 2870 $\pm$ 27 & 1.363 $\pm$ 0.048 & 0.257 $\pm$ 0.074 & 38 $\pm$ 3 & 5.35 & 0.23 $\pm$ 0.01 & 0.04104 & 0.343 \\
14891 & 4928 $\pm$ 89 & 1.121 $\pm$ 0.055 & 0.331 $\pm$ 0.060 & 167 $\pm$ 8 & 2.62 & -0.06 $\pm$ 0.06 & 0.00774 & 0.174 \\
20627 & 1180 $\pm$ 12 & 0.273 $\pm$ 0.016 & 0.345 $\pm$ 0.049 & 90 $\pm$ 17 & 2.68 & 0.14 $\pm$ 0.02 & 0.00194 & 0.103 \\
\hline       
\end{tabular}
\end{table*}

Using the mass function, we can estimate the minimum mass ($M_{\rm \ast}$) of the companion, under certain assumption on the mass of the RRL component \citep[$M_{\rm RRL}$ = 0.65 M$_{\odot}$,][]{Prudil2019OO} and inclination angle $i$ = 90\,deg:
\begin{equation}
f(\mathfrak{M}) = \frac{M_{\rm \ast}^{3} \cdot \text{sin}^{3}\,i}{\left(M_{\rm RRL} + M_{\rm \ast} \right)^{2} }.
\end{equation}
We note that a difference of 0.1 M$_{\odot}$ in the assumed mass of the RRL variable only marginally affects the estimated minimum mass (less than 0.07\,M$_{\odot}$ for a companion with the highest mass function). High uncertainty is in the unknown inclination angle. 

\begin{figure*}
\includegraphics[width=2\columnwidth]{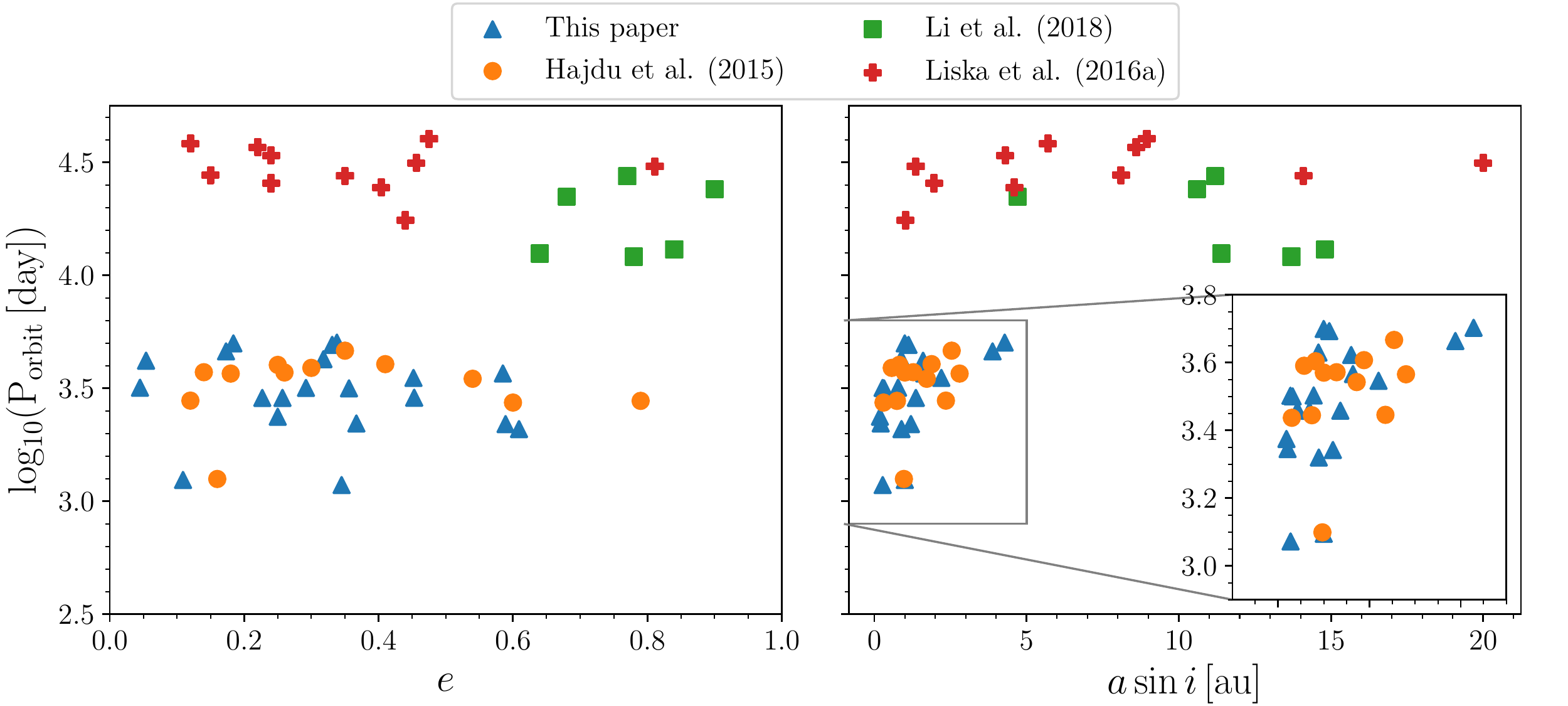}
\caption{Comparison of orbital parameters for binary candidates in our study (blue triangles) and studies by \citet[][orange points]{Hajdu2015}, \citet[][red plus signs]{Liska2016b}, and \citet[][green squares]{Li2018}. In both panels, we show the orbital period on the horizontal axes, and on the vertical axes eccentricity (left-hand panel), and semi-major axis (right-hand panel).}
\label{fig:OrbitalParameters}
\end{figure*}

In Fig.~\ref{fig:OrbitalParameters} we show a comparison of orbital parameters from our study and from studies by \citet{Hajdu2015,Liska2016b,Li2018}. We compare orbital periods, eccentricity, and semi-major axes of the candidates from the aforementioned studies. In these two panels, we see a gap in the orbital periods between our study and the study by \citet{Hajdu2015}, and the orbital period from \citet{Liska2016b,Li2018}. This is most likely due to the limited time span of the OGLE GB survey, which reaches up to 25 years while the studies by \cite{Liska2016b} and \cite{Li2018} consist of stars with more than a hundred years of observation. The eccentricity seems to be independent of the orbital period even though we detected a small correlation during the $O-C$ modeling for some of our sample stars. The connection between the semi-major axis and orbital period comes from the third Kepler law.  We note that we do not detect any evident connection between the orbital parameters and pulsation periods.

\section{Discussion and Conclusion} \label{sec:Discus}

In our study, we found 20 binary candidates among more than 9000 non-modulated RRL stars from the Galactic bulge. The low number of candidates further supports a low occurrence of RRL variables in binary systems. Regarding a possible confirmation of binary candidates, we rely mainly on the orbital periods and semi-amplitude of the radial velocity of RRL component orbit, $K$, which can highlight feasible candidates for spectroscopic follow-up. 

The orbital periods range from $\approx$ 3 years up to 14 years. This is comparable with the study by \citet{Hajdu2015} of a similar dataset, but when compared with the studies by \cite{Liska2016b} and \cite{Li2018}, our candidates lie on the short end of the orbital period distribution. The semi-amplitudes of the radial velocity for our candidates lie on a boundary for detection in studies of radial velocities among the bulge RRL stars \citep[][radial velocities with errors between 5-10\,km\,s$^{-1}$]{Kunder2016}. Therefore, with current instruments, it is possible to follow-up and possibly confirm binary candidates as shown in \citet{Hajdu2018}, although several years of spectroscopic observations are a necessity. 

Suitable candidates for such spectroscopic radial-velocity follow-up are e.g. OGLE-BLG-RRLYR-02854, 07051, and 07640 with a rather high semi-amplitude of the radial velocity (above 7\,km\,s$^{-1}$) and rather short orbital periods (less than 10\,years). All three RRL stars should have a companion with a minimum mass close to the assumed mass of the RRL component. Two of the binary candidates from our sample (OGLE-BLG-RRLYR-08185 and 13477) should have a companion that is more massive than the RRL component. If the more  massive companion is at the asymptotic giant branch, it can contribute to the total brightness of the system, and with a sufficiently long major axis with appropriate inclination might be detectable using the {\it Gaia} space telescope \citep{GAIA2016,GaiaBrown2018}. In addition, anomalies in proper motions can help to identify potential candidates \citep{Kervella2019a,Kervella2019b}.

The OGLE survey has been proven in the past to suffice for a binarity search among RRL stars, but in order to confirm the found candidates as binary systems, the combination of spectroscopic measurements with photometry is obligatory.

\section*{Acknowledgements}
This research has made use of the KMTNet system operated by the Korea Astronomy and Space Science Institute (KASI) and the data were obtained at three host sites of CTIO in Chile, SAAO in South Africa, and SSO in Australia. Z.P. acknowledges the support of the Hector Fellow Academy. M.S. acknowledges support from Postdoc$@$MUNI project CZ.02.2.69/0.0/0.0/16$\_$027/0008360. This research was carried out under the project CEITEC 2020 (LQ1601) with financial support from the Ministry of Education, Youth and Sports of the Czech Republic under the National Sustainability Programme II. E.K.G was supported by Sonderforschungsbereich SFB 881 "The Milky Way System" (subprojects A03, A11) of the German Research Foundation (DFG).  Work by C.-U. L. was supported by KASI (Korea Astronomy and Space Science Institute) grant 2018-1-830-02.


\bibliographystyle{mnras}
\bibliography{biby-old}

\newpage 


\bsp	
\label{lastpage}
\end{document}